\documentclass[aps,pra,reprint,superscriptaddress,showpacs,showkeys]{revtex4-1}

\usepackage[T1]{fontenc}
\usepackage{graphicx}
\usepackage{amsmath}
\usepackage{textcomp}

\begin{document}


\title{Absolute strong-field ionization probabilities of ultracold rubidium atoms}


\author{Philipp Wessels}
\email[]{pwessels@physnet.uni-hamburg.de}
\homepage[]{http://www.cui.uni-hamburg.de/}
\affiliation{The Hamburg Centre for Ultrafast Imaging, Luruper Chaussee 149, 22761 Hamburg, Germany}
\affiliation{Center for Optical Quantum Technologies, University of Hamburg, Luruper Chaussee 149, 22761 Hamburg, Germany}

\author{Bernhard Ruff}
\affiliation{The Hamburg Centre for Ultrafast Imaging, Luruper Chaussee 149, 22761 Hamburg, Germany}
\affiliation{Center for Optical Quantum Technologies, University of Hamburg, Luruper Chaussee 149, 22761 Hamburg, Germany}

\author{Tobias Kroker}
\affiliation{The Hamburg Centre for Ultrafast Imaging, Luruper Chaussee 149, 22761 Hamburg, Germany}
\affiliation{Center for Optical Quantum Technologies, University of Hamburg, Luruper Chaussee 149, 22761 Hamburg, Germany}

\author{Andrey K. Kazansky}
\affiliation{Departamento de Fisica de Materiales, UPV/EHU, 20018 San Sebastian/Donostia, Spain}
\affiliation{Ikerbasque, Basque Foundation for Science, 48011 Bilbao, Spain}
\affiliation{Donostia International Physics Center (DIPC), 20018 San Sebastian/Donostia, Spain}

\author{Nikolay M. Kabachnik}
\affiliation{The Hamburg Centre for Ultrafast Imaging, Luruper Chaussee 149, 22761 Hamburg, Germany}
\affiliation{Donostia International Physics Center (DIPC), 20018 San Sebastian/Donostia, Spain}
\affiliation{Skobeltsyn Institute of Nuclear Physics, Lomonosov Moscow State University, Moscow 119991, Russia}

\author{Klaus Sengstock}
\affiliation{The Hamburg Centre for Ultrafast Imaging, Luruper Chaussee 149, 22761 Hamburg, Germany}
\affiliation{Center for Optical Quantum Technologies, University of Hamburg, Luruper Chaussee 149, 22761 Hamburg, Germany}

\author{Markus Drescher}
\affiliation{The Hamburg Centre for Ultrafast Imaging, Luruper Chaussee 149, 22761 Hamburg, Germany}
\affiliation{Center for Optical Quantum Technologies, University of Hamburg, Luruper Chaussee 149, 22761 Hamburg, Germany}

\author{Juliette Simonet}
\affiliation{Center for Optical Quantum Technologies, University of Hamburg, Luruper Chaussee 149, 22761 Hamburg, Germany}


\date{February 21, 2018}

\begin{abstract}
 We report on precise measurements of absolute nonlinear ionization probabilities obtained by exposing optically trapped ultracold rubidium atoms to the field of an ultrashort laser pulse in the intensity range of $1 \times 10^{11}$ to $4 \times 10^{13}$~W/cm$^2$. The experimental data are in perfect agreement with \emph{ab-initio} theory, based on solving the time-dependent Schr\"odinger equation without any free parameters. Ultracold targets allow to retrieve absolute probabilities since ionized atoms become apparent as a local vacancy imprinted into the target density, which is recorded simultaneously. We study the strong-field response of $^{87}$Rb atoms at two different wavelengths representing non-resonant and resonant processes in the demanding regime where the Keldysh parameter is close to unity.
\end{abstract}


\maketitle


Ultracold atoms serve as ideal systems for precise studies of light-matter interaction with a high degree of control over critical parameters. The creation of charged particles in form of ions and electrons out of these well-defined ensembles offers appealing possibilities for fundamental research as well as advanced applications. Hybrid atom-ion quantum systems benefit from long-range interactions introduced by the Coulomb potential \cite{Tomza:ColdAtomIonReview, Haerter:ColdAtomIonReview} and are promising candidates for a quantum information platform \cite{Doerk:AtomIonQG, Idziaszek:AtomIonCollisions}. Moreover, ions immersed in a neutral ultracold ensemble can form new states of matter such as mesoscopic molecular ions \cite{Schurer:MesoscopicIon, Schurer:IonCapture} and ultracold plasma \cite{Killian:Plasma, Killian:PlasmaReview, Killian:UNPPhysReports}.

Ultrashort laser pulses of durations in the femtosecond domain offer a mechanism for instantaneous creation of cold ions and electrons in ultracold environments with respect to the time-scales of the atomic motion in the trap. Experiments combining ultracold targets with ultrashort laser pulses are able to observe detailed space-charge dynamics \cite{Murphy:RbIonsSC, Takei:ManyBodyElDynamics} and provide ultracold electron bunches exhibiting a large degree of coherence for the use in high-brilliance accelerator sources \cite{Claessens:UltracoldElectronSource, McCulloch:ElectronBunches}. The ionization is achieved at high peak intensities of the laser field either through absorption of multiple photons or a substantial modification of the Coulomb potential that leads to tunneling \cite{Protopapas:HighIntensityReview, Agostini:IonizationStrongFieldReview}.

For all of the above mentioned examples of cold atom/ion and electron systems it is vital to understand the nature of the ionization process in detail since the subsequent dynamics strongly depend on the ionization regime. The high degree of controllability on the atomic side eventually has to be extended also to the process of creating electrons and ions on instantaneous time-scales with ultrashort laser pulses.

Traditionally, the ionization of atoms in strong light fields is categorized into intuitive regimes distinguished by the Keldysh adiabaticity parameter \cite{Keldysh:Ionization}

\begin{equation}
 \gamma = \sqrt{\frac{E_I}{2 U_p}} \qquad \text{with} \qquad U_p = \frac{e^2 I}{2 c \varepsilon_0 m_e \omega^2}
 \label{eq:keldysh}
\end{equation}

\noindent relating the ionization potential $E_I$ to the average quiver energy of a free electron with mass $m_e$ and charge $e$. The latter is described by the ponderomotive potential $U_p$ generated by the driving laser field of angular frequency $\omega$ and intensity $I$. A perturbative multiphoton model ($\gamma \gg 1$) is contrasted by an adiabatic tunnel ionization \cite{PPT:Ionization, ADK:Tunnel, Augst:TunnelIonization} regime ($\gamma \ll 1$) that leads to over-the-barrier ionization for increasing laser intensities \cite{Protopapas:HighIntensityReview, Agostini:IonizationStrongFieldReview}.

So far, the ionization process in strong laser fields has been studied mainly with rare gas atoms. For alkali atoms used in most cold atom experiments the transition regime separating multiphoton from tunnel ionization at a Keldysh parameter around $\gamma \approx 1$ is reached at considerably lower intensities. Experiments accessing the strong-field regime with alkali atoms \cite{Schuricke:StrongFieldLi, Schuricke:LiCoherence, Hart:SFIBSINa} have revealed that commonly used strong-field ionization models in form of a pure multiphoton- or tunnel ionization process can not be applied \emph{a priori} which makes employing \emph{ab-initio} calculations by solving the time-dependent Schr\"odinger equation (TDSE) inevitable.

An accurate comparison of theoretical models to experimental data sets is notoriously difficult since the ionization of atoms in strong laser fields depends nonlinearly on a variety of experimental parameters which are not always accessible or under control. Often, the intensity calibration is even performed using the theory under test, for example in advanced electron spectroscopy experiments \cite{Boge:AttoclockSFI, Hofmann:NonAdiabaticSF}.

Measuring absolute ionization probabilities is in general even more challenging due to the necessity of precise knowledge on the incoming photon-flux, target density as well as detector efficiencies. Ionization cross-sections can be accurately measured using magneto-optically trapped (MOT) atoms and observing the trap loss rates before and after introducing an ionizing laser beam \cite{Dinneen:TrapLoss, Madsen:AbsoluteMgCS, Anderlini:TwoPhotonColdRb}. This technique is very sensitive and can measure low ionization rates. However, it relies on calibrating the excited state fraction because of the intrinsic presence of MOT laser beams. Additionally, a focused laser beam always consists of a distribution of intensities, a phenomenon usually referred to as focal averaging that has to be taken into account.

We present an alternative methodology to overcome these obstacles by employing optically trapped, ultracold $^{87}$Rb ensembles at nanokelvin temperatures. We observe local losses imprinted as vacancies into the simultaneously detectable atomic density profile which leads to a spatially resolved loss signal. This allows us to extract absolute ionization probabilities. We employ this technique to precisely measure strong-field losses in $^{87}$Rb atoms and compare the measurements to the results obtained using ionization probabilities calculated by numerically solving the TDSE. We find an excellent agreement between experiment and theory \emph{without any free parameters}. The sole outer-shell electron in alkali atoms establishes a reference system for studying strong-field ionization in the transitional $\gamma \approx 1$ regime.

\section*{Results}
\label{sec:results}

\subsection*{Measuring absolute local losses}
\label{sec:methodology}

\begin{figure}
 \includegraphics[width=1\linewidth]{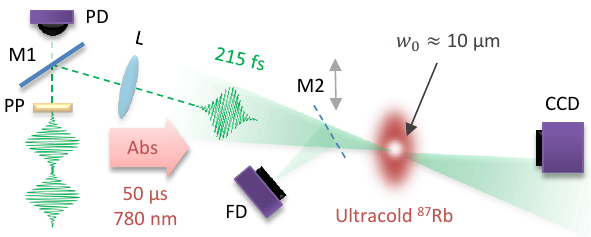}%
 \caption{\label{fig:setup} Probing strong-field ionization with ultracold $^{87}$Rb atoms. A single femtosecond laser pulse selected by a pulse picker (PP) is focused by a $f = 200$~mm lens (L) to a waist of $w_0 \approx 10$~\textmu m onto a cloud of $^{87}$Rb atoms in an optical dipole trap. Ionization losses in a known density distribution are measured using resonant absorption imaging. The leakage through a mirror (M1) is used for simultaneous measurement of the pulse energy on a photo diode (PD) and the focus featuring different areas of intensities is imaged onto a diagnostics camera (FD) through a moveable mirror (M2).}
\end{figure}

In order to achieve nanokelvin temperatures, $^{87}$Rb captured and laser-cooled in a 2D/3D MOT is further cooled by transferring the atoms into a hybrid trap \cite{Lin:HybridTrap} which consists of a magnetic quadrupole field and a crossed optical dipole trap at 1064~nm wavelength. This realizes a well defined sample of ground state $^{87}$Rb atoms at final temperatures around 100~nK close to quantum degeneracy (for details see Methods section). As presented in Fig.~\ref{fig:setup}, a single femtosecond laser pulse delivered by a pulse-picker is focused onto the trapped atoms where it locally ionizes $^{87}$Rb atoms in the cloud. The remaining atomic density is mapped by resonant absorption imaging at the $^{87}$Rb $D_2$ transition (780~nm wavelength) by a 50~\textmu s long imaging pulse. The focal intensity distribution is imaged and the pulse energy is recorded simultaneously to accurately characterize the intensity in the focal volume for calibrated peak intensities.

This experiment is performed at two different wavelengths representing non-resonant and resonant ionization processes. The non-resonant case is realized at a wavelength of 511~nm with pulses of $215^{+20}_{-15}$~fs FWHM duration while for the resonant case pulses of 1022~nm wavelength are used at a pulse duration of $300^{+30}_{-23}$~fs FWHM. In both cases the femtosecond laser beam is focused to a waist of $w_0 \approx 10$~\textmu m using a lens of $f = 200$~mm focal length (for exact numbers and details see Methods section).

\subsection*{Data evaluation and comparison to theory}
\label{sec:dataevaluation}

\begin{figure}
 \includegraphics[width=1\linewidth]{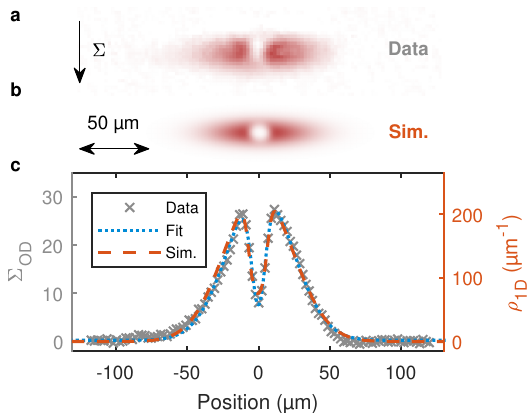}%
 \caption{\label{fig:clouds} Quantifying ionization losses in an ultracold atomic ensemble. \textbf{a} Measured and \textbf{b} calculated in-situ density distribution of ultracold $^{87}$Rb atoms after applying a focused femtosecond laser pulse of 511~nm wavelength and $6.9^{+0.7}_{-0.8} \times 10^{12}$~W/cm$^2$ peak intensity. \textbf{c} The row sum profiles for the measured (gray crosses) and simulated (red dashed line) density images are used to quantify the measured and simulated fraction of ionized atoms by applying a double Gaussian fit (blue dotted line for measured profile).}
\end{figure}

Figure~\ref{fig:clouds}a shows an in-situ absorption image of the atomic density exposed to a single femtosecond laser pulse of $I_0 = 6.9^{+0.7}_{-0.8} \times 10^{12}$ W/cm$^2$ peak intensity. The ionization losses lead to a vacancy imprinted into the atomic density profile. In contrast to typical MOT temperatures of approximately 100~\textmu K, here the motion of atoms is negligible during imaging so that we have access to the local losses in the focal intensity distribution and simultaneously measure the line-of-sight integrated two-dimensional density profile of the neutral atoms. Most other experiments are limited to an integrated signal over the focal volume as sole observable and can not access the local density profile. In order to quantify the fraction of lost atoms, the difference of two Gaussian functions

\begin{eqnarray}
 \Sigma_\text{OD}(x) = A_0 + A_c \exp{\left(- \frac{\left(x - x_c\right)^2}{2 \sigma_c^2} \right)}\nonumber\\
                           - A_v \exp{\left(- \frac{\left(x - x_v\right)^2}{2 \sigma_v^2} \right)}%
 \label{eq:doublegaussian}
\end{eqnarray}

\noindent is fitted to the pixel row sum of the image in Fig.~\ref{fig:clouds}c where $A_0$ determines an offset while $A_c$, $x_c$ and $\sigma_c$ correspond to amplitude, central position and width of the atomic cloud and $A_v$, $x_v$ and $\sigma_v$ describe the same parameters for the vacancy, respectively. The loss fraction

\begin{equation}
 f_l = \frac{N_i}{N_0} = \frac{A_v \sigma_v}{A_c \sigma_c}%
 \label{eq:fl}
\end{equation}

\noindent relating the number of ionized atoms $N_i$ to the total initial number of atoms $N_0$ can be derived by the fit parameters as shown in Eq.~(\ref{eq:fl}).

In order to compare the measurements to theoretically obtained ionization probabilities, the predicted loss-fraction due to photoionization is calculated. First, the probability $P$ of an atom to be ionized after the laser pulse of peak intensity $I_0$ is derived from numerical TDSE calculations. Using these results, the distribution of peak intensities within the laser focus $I_0(x,y,z)$ is mapped to spatially resolved ionization probabilities $P(x,y,z)$ which are superimposed with the 3D density map of the atomic sample for each grid unit (for details see Methods section). This results in a simulated 3D density distribution of the remaining, non-ionized atoms that is projected onto the image plane of the camera and convoluted with the imaging resolution as shown in Fig.~\ref{fig:clouds}b. The resulting simulated 2D projection is evaluated in the same manner as the absorption image. A remarkable agreement is obtained between the calculated and measured row sum as depicted in Fig.~\ref{fig:clouds}c. The simulated loss fraction $f_l^\text{sim}$ can be extracted from the fit to the computed profile and compared to the experimentally obtained value.

\subsection*{Non-resonant two-photon ionization at 511~nm}
\label{sec:511nm}

\begin{figure}
 \includegraphics[width=1\linewidth]{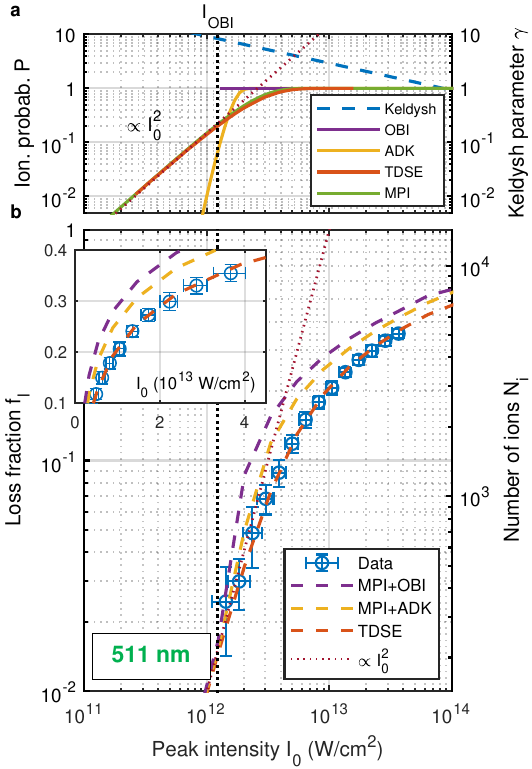}%
 \caption{\label{fig:data_511nm} Non-resonant strong-field ionization yield of ultracold $^{87}$Rb. \textbf{a} Keldysh parameter $\gamma$ and calculated ionization probabilities $P$ after a single laser pulse for different ionization models and \textbf{b} measured fraction of lost atoms $f_l$ (blue dots) for laser pulses of 511 nm wavelength. The nominal over-the-barrier intensity $I_\text{OBI}$ is depicted as vertical black dashed line. The data is compared to the calculated loss fraction $f_l^\text{sim}$ using \emph{ab-initio} TDSE calculations (red) and a two-photon ionization cross section of $\sigma_2 = 1.5 \times 10^{-49}$ cm$^4$ s \cite{Takekoshi:RbMPI} (green). The red dotted lines indicate the $I_0^2$ scaling of a two-photon process. The additional assumption in the calculation that the electron is free once $I_\text{OBI}$ is exceeded leads to an overestimation of the losses (purple lines) just as including tunnel-ionization losses using pure ADK theory (yellow lines). The inset in \textbf{b} presents the same data in linear instead of log/log scaling.}
\end{figure}

The measured fraction of lost atoms $f_l$ according to Eq.~(\ref{eq:fl}) with respect to the peak intensity $I_0$ is collected in Fig.~\ref{fig:data_511nm}b for 511~nm wavelength. The corresponding number of generated ions $N_i$ is computed by multiplying the fraction of lost atoms with the initial number of atoms in the cloud and is depicted on the right axis. The increasing measured loss fraction (blue dots) originates from higher ionization probabilities $P$ at higher peak intensities presented in Fig.~\ref{fig:data_511nm}a. Once the ionization probability is close to unity in the focal center (here at $5 \times 10^{12}$~W/cm$^2$), losses still increase since the volume of high ionization probabilities expands over the atomic cloud.

The calculated loss fraction $f_l^\text{sim}$ expected by strong-field ionization probabilities obtained from \emph{ab-initio} TDSE calculations is shown as red dashed line in Fig.~\ref{fig:data_511nm}b and perfectly reproduces the experimental result. It should be emphasized that this curve neither has free parameters to fit to the data points, nor is it scaled or shifted so that the absolute ionization yield of the focal intensity distribution can be directly compared to the measurement because we image local losses in a simultaneously observable atomic density profile.

Although the TDSE method uses no other approximation than the single active electron approximation, it is still helpful to compare our measurements to widely used strong-field ionization models to gain intuition regarding the applicability of these models as well as the sensitivity of our measurements to deviations from the TDSE ionization probabilities. For these reasons we now compare our datasets to common strong-field ionization models.

In a multiphoton ionization process at 511~nm wavelength, the rubidium atom has to absorb two photons. For reference, the expected ionization probability described by lowest order perturbation theory using a multiphoton ionization (MPI) cross-section of $\sigma_2 = 1.5 \times 10^{-49}$~cm$^4$~s at the given wavelength \cite{Takekoshi:RbMPI} is included in Fig.~\ref{fig:data_511nm}a as green solid line. This model agrees surprisingly well with the \emph{ab-initio} results over the whole intensity range even though the ionization probability as well as the Keldysh parameter are close to unity.

In contrast to rare gas atoms, the electric field that classically suppresses the Coulomb barrier of the atom to free the valence electron is exceeded at much lower intensities in alkali atoms due to the low binding energy. For rubidium this over-the-barrier (OBI) intensity is reached at $I_\text{OBI} = 1.22 \times 10^{12}$~W/cm$^2$ (vertical black dashed line). Previous experiments \cite{Schuricke:StrongFieldLi, Hart:SFIBSINa} and theoretical studies \cite{Morishita:LiSFTheory, Jheng:LiIonTheory} have already mentioned the peculiar situation that in alkali atoms the OBI intensity is reached well within the multiphoton regime, in particular for rubidium at a Keldysh parameter of $\gamma = 8.4$ for an optical wavelength of 511~nm (compare Fig.~\ref{fig:data_511nm}a). Assuming that the ionization probability is unity once the OBI intensity is exceeded \cite{DK:TunnelBSI} and is given by the multiphoton ionization cross section otherwise, we can test for contributions of over-the-barrier ionization processes. The expected loss fraction in this case significantly overestimates the observed loss fraction (purple lines in Fig.~\ref{fig:data_511nm}).

Although the Keldysh parameter is still bigger than (but close to) unity, we can also test the tunnel-ionization model by calculating ionization probabilities obtained by the Ammosov-Delone-Krainov (ADK) theory \cite{ADK:Tunnel} that is strictly valid only for Keldysh parameters much smaller than unity ($\gamma \ll 1$). Details on this calculations are given in the Supplemental Material. Including this loss channel into the calculation (yellow lines in Fig.~\ref{fig:data_511nm}) still clearly overestimates the observed ionization yield. Regarding the sensitivity of our method it should be noted that the small intensity range from $1.5$ to $4 \times 10^{12}$~W/cm$^2$ where the ADK ionization probability exceeds the MPI probability, leads to an experimentally expected loss fraction that lies significantly outside the error bars of the measurements so that our technique is sensitive to even subtle details in the ionization probability curve.

We observe considerable deviations from all conventional adiabatic models which is expected from a Keldysh parameter still bigger than unity. This means that the modification of the Coulomb barrier of the atom \emph{cannot} be regarded as adiabatic which also renders the concepts of an over-the-barrier ionization intensity not applicable in this case. Consequently, one has to rely on calculating the ionization process with TDSE methods. Our results underline the dominance of multiphoton ionization in alkali atoms consistent with previous experiments \cite{Schuricke:StrongFieldLi, Schuricke:LiCoherence, Hart:SFIBSINa} and theory \cite{Morishita:LiSFTheory, Jheng:LiIonTheory}.

\subsection*{Resonant multiphoton ionization at 1022~nm}
\label{sec:1022nm}

\begin{figure}
 \includegraphics[width=1\linewidth]{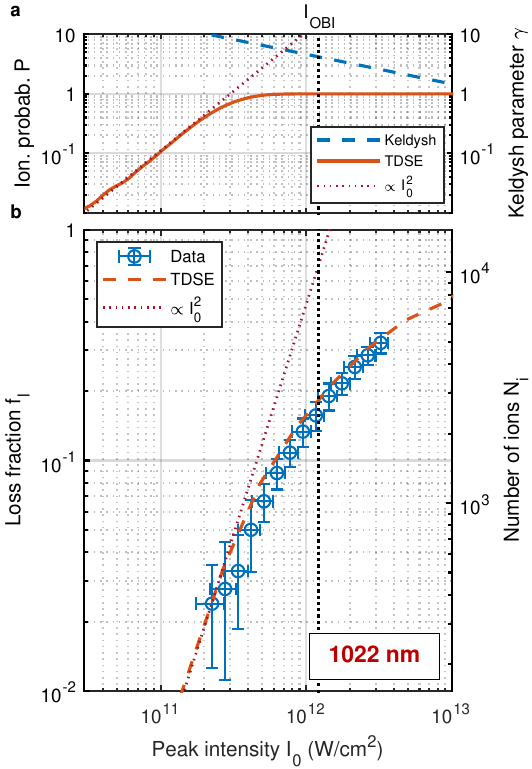}%
 \caption{\label{fig:data_1022nm} Resonant strong-field ionization yield of ultracold $^{87}$Rb. \textbf{a} Keldysh parameter $\gamma$ (blue dashed line) and ionization probabilities $P$ after a single laser pulse obtained by \emph{ab-initio} TDSE calculations (red solid line). \textbf{b} Measured fraction of lost atoms $f_l$ (blue dots) for laser pulses of 1022~nm wavelength and comparison to the loss fraction $f_l^\text{sim}$ resulting from TDSE calculations (red dashed line). The nominal over-the-barrier intensity $I_\text{OBI}$ is depicted as vertical black dashed line. The red dotted lines indicate an $I_0^2$ scaling.}
\end{figure}

Figure~\ref{fig:data_1022nm} covers the resonant ionization at smaller photon energies using pulses of 1022~nm wavelength. Compared to the measurement at 511~nm in Fig.~\ref{fig:data_511nm}, significant losses become evident at lower intensities. The calculated ionization probabilities obtained by TDSE calculations are presented in Fig.~\ref{fig:data_1022nm}a. In the absence of resonances, a multiphoton ionization process at this wavelength involves four photons and would scale with $I_0^4$. For comparison, an $I_0^2$ scaling is plotted along with the theoretical results as the most compatible scaling law with the calculated ionization probability curve.

The experimental data points in Fig.~\ref{fig:data_1022nm}b also agree extremely well with the expected loss fraction from the TDSE calculations in view of the fact that there is no free parameter and no arbitrary scaling involved. Compared to the resonant measurement the focal intensity distribution here was not perfectly Gaussian. Figure~\ref{fig:foci} depicts the measured foci for both wavelengths obtained by the focal diagnostics (see also Fig.~\ref{fig:setup}). While the focus of the 511~nm laser pulse is well represented by a single 2D Gaussian fit, the focus image at 1022~nm wavelength features additional weak intensity regions next to the central main intensity peak. In the calculations of the loss fraction with the TDSE ionization probabilities we have accounted for this by approximating the measured intensity distribution by a superposition of three Gaussian profiles. The reconstruction is discussed in the Methods section and also presented in Fig.~\ref{fig:foci}. The remaining minor discrepancy between calculated and measured loss fraction is believed to be resolved by a more accurate intensity model. The larger experimental vertical error bars compared to the 511~nm results are due to less statistics collected in this measurement.

\begin{figure}
 \includegraphics[width=1\linewidth]{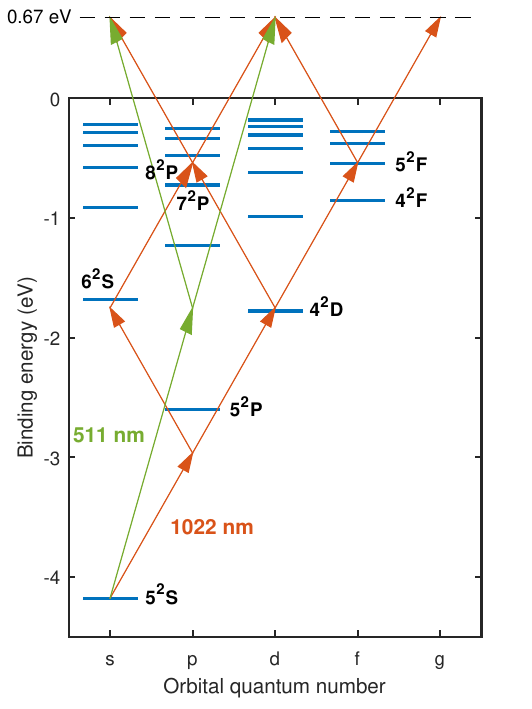}%
 \caption{\label{fig:rb_levels} $^{87}$Rb energy levels and multiphoton transitions. The green arrows represent photons at 511~nm wavelength while the red arrows represent photons at 1022~nm wavelength. The excess energy of 0.67~eV transferred to the released electron is indicated as black dashed line. In a four-photon transition the $4^2D$ and $5^2F$ states are in reach for a resonant ionization process.}
\end{figure}

The resonant $I_0^2$ scaling at 1022~nm can be explained using multiphoton transition pathways and energy levels of $^{87}$Rb depicted in Fig.~\ref{fig:rb_levels}. The unperturbed energy levels are experimental NIST values \cite{Safronova:Rubidium}. This figure indicates that the apparent high ionization probability can be induced by the $4^2D$ and $5^2F$ states. The $4^2D$ resonance is energetically separated from the $5^2S$ ground state by 2.40~eV which is close to the 2.43~eV energy difference corresponding to two 1022~nm photons but just outside the spectral bandwidth of the laser pulse. The $5^2F$ resonance, however, is separated from the ground state by 3.63~eV so that it is accessible using three photons which add up to an energy of 3.64~eV with 0.02~eV bandwidth (see Methods for details).

\begin{figure}
 \includegraphics[width=1\linewidth]{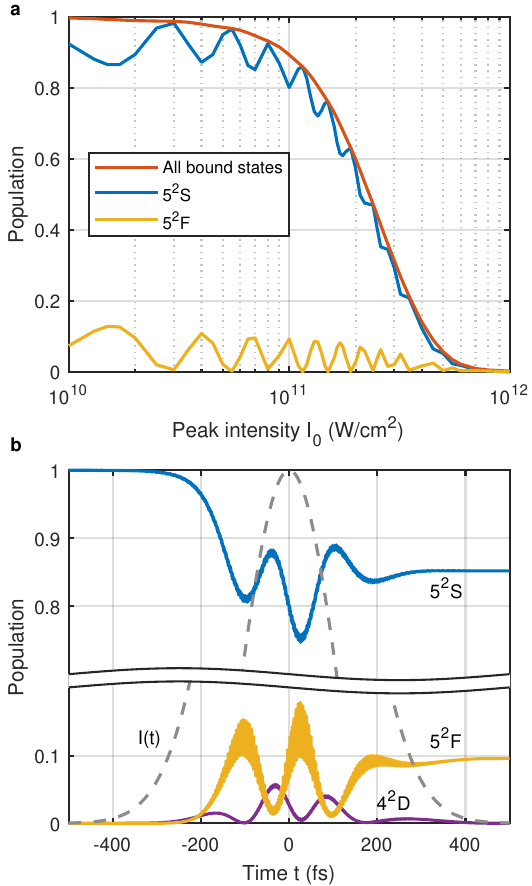}%
 \caption{\label{fig:rb_population} Population of bound states. \textbf{a} The integral population of all $^{87}$Rb bound states after a Gaussian laser pulse of 1022~nm wavelength and 300~fs FWHM duration is shown with respect to the peak intensity $I_0$ (red solid line). Population of the $5^2S$ ground state (blue) and $5^2F$ resonant state (yellow) shows that after the laser pulse the population is mainly transferred from the ground state into the continuum and into the $5^2F$ state. The final $5^2F$ population oscillates with the peak intensity. \textbf{b} Time-dependent population of the $5^2S$ ground state (blue) as well as the $5^2F$ (yellow) and $4^2D$ (purple) excited states during the laser pulse of $I_0 = 7 \times 10^{10}$~W/cm$^2$ peak intensity (gray dashed line).}
\end{figure}

By exposing the atom to the femtosecond laser beam, the atomic energy levels are modified by a time-dependent AC Stark shift during the laser pulse so that bound states can shift into or out of resonance. In order to answer the question whether the resonant process is a $2 + 1 + 1$ photon process involving both, the $4^2D$ and $5^2F$ resonances or rather a $3 + 1$ photon process via the $5^2F$ state we present further results from the TDSE calculations regarding the population of bound states after the laser pulse in Fig.~\ref{fig:rb_population}a. By projecting the electron wave packet to the $5^2S$ ground state and to the $4^2D$ or $5^2F$ excited state, respectively, we can trace the final population of these states with respect to the intensity.

Additionally, we show the sum over all bound states of the atom. This survival fraction is used to derive the ionization probability by subtracting it from unity. It is apparent that the depopulation of the ground state after the pulse results in a final population of the $5^2F$ state and ionization whereas the contribution of the $4^2D$ state is negligible. The oscillations of the population of the ground and $5^2F$ states presumably are due to Rabi-type oscillations in a complex system of three bound states and the continuum.

Because bound states are clearly populated and we experimentally detect remaining, non-ionized atoms on the $^{87}$Rb $D_2$ transition, one could argue that the experiment only detects non-ionized atoms if they are in the $5^2S$ ground state. However, if solely the $5^2F$ excited state is populated in addition to the ground state, a single photon of the 780~nm detection pulse can also be absorbed on the $5^2F$ to continuum transition which counts as a non-ionized atom in the absorption imaging. This is why the ionization probability here is derived from the sum over all bound states in Fig.~\ref{fig:rb_population}a.

Figure~\ref{fig:rb_population}b shows the population of bound states during a laser pulse of $I_0 = 7 \times 10^{10}$~W/cm$^2$ peak intensity. Although the $4^2D$ resonance is negligibly populated after the pulse, a transient population during the pulse can be observed which enhances the ionization probability. The plot shows an oscillatory population transfer with a frequency of approximately 7~THz which modulates much faster oscillations not resolved in the figure. A Fourier analysis of the fast oscillations shows that the $4^2D$ and $5^2F$ populations follow a signal corresponding to mainly twice the laser frequency. This is an indication for a two-photon population transfer, thus explaining the $I_0^2$ scaling. Similar theoretical studies have also reported such oscillations of final states with respect to the intensity and time for alkali atoms \cite{Morishita:LiSFTheory, Jheng:LiIonTheory, Jheng:LiWavePacketDisplacement}.

\section*{Conclusion and outlook}
\label{sec:conclusion_outlook}

In summary, by cooling the atomic system considerably below MOT temperatures to approximately 100~nK in an optical far-off resonance trap we image local losses imprinted into the measurable atomic density profile. We have experimentally determined absolute ionization yields in ultracold atomic ensembles and have extracted absolute ionization probabilities for non-resonant and resonant ionization processes at two different wavelengths. We provide accurate data sets for precise tests and compare our experimental observation, without any arbitrary scaling, to ionization probabilities obtained by numerically solving the TDSE.

Our methodology allows for a pixel-wise evaluation of the losses for truly overcoming focal-averaging and measuring ionization processes at one single intensity. This would provide a simultaneous measurement at several intensities in the focal intensity distribution and will be highly beneficial for sources where parameters such as pulse duration and pulse energy fluctuate. Moreover, this would give a more direct comparison to theory where single intensities are used for calculations. Additionally, by varying the wavelength of the detection pulse, sensitive tracing of intermediate state population after the laser pulse is possible. In principle, the aforementioned methods can also be applied in a time-resolved manner by employing pump-probe techniques.

In contrast to magneto-optical traps, atoms trapped optically by far-off resonant laser light have the advantage of a vanishing excited state fractions and the absence of electric or magnetic fields. This is beneficial for electron or ion spectroscopy after ionization \cite{Sharma:AllOpticalMOT} which can reveal further insight into strong-field physics in alkali atoms. Our experimental techniques can also be used for absolute calibration of detector efficiencies in the rapidly growing number of experiments using cold atoms with charged particle detection \cite{Henkel:SingleIonDetection, Stecker:ColdIonMicroscope, McCulloch:ElectronBunches, Claessens:UltracoldElectronSource, Takei:ManyBodyElDynamics, Goetz:UltracoldIonizationLED}.

Recent publications discuss that ponderomotive shifts for alkali atoms substantially lower the ionization probability compared to direct ADK results due to the high polarizability of alkali atoms \cite{Milosevic:StrongFieldAlkali, Bunjac:SFQCSodium, Miladinovic:AlkaliBSI}. When using longer pulsed laser wavelengths where the adiabatic approximation is genuinely valid, these effects have to be taken into account and can be tested with a high sensitivity with our method. Note, however, that ADK does not take into account intermediate resonances which become important for longer wavelengths. Moreover, the population of bound states after the laser pulse and the critical influence of an AC Stark shift can impose coherent control schemes for excitations in ultracold samples \cite{Krug:CoherentStrongFieldControl}.

Particularly interesting experiments include the investigation of the strong-field ionization of ultracold atoms before and after the Bose-Einstein condensation (BEC) phase transition. Previous results have been inconclusive whether the ionization rate increases or decreases in this case \cite{Ciampini:BECPhotoionization, Mazets:BECIonizationSuppression}.

Finally, the presented measurements reveal the strong-field response of alkali atoms by accessing a transitional regime where the Keldysh parameter and the ionization probability are close to unity. The single outer-shell electron and the low ionization potential combined with a high polarizability compared to commonly used rare gas atoms make ultracold alkali atoms ideal model systems for studying strong-field physics.

\footnotesize

\section*{Methods}

\subsection*{Ultracold rubidium ensembles}
\label{mth:ultracold}

Rubidium atoms are thermally evaporated from a dispenser into a vacuum system where the $^{87}$Rb isotope is captured in a 2D MOT and transferred into a 3D MOT by a near-resonant pushing laser beam. After the 12~s long MOT loading phase, the atoms are cooled in a 4 ms expansion during an optical molasses phase. Subsequently, the atoms are transferred into a hybrid trap \cite{Lin:HybridTrap}. This trap consists of a magnetic quadrupole field generated by a pair of coils in anti-Helmholz configuration and a far-detuned crossed optical dipole trap at 1064~nm wavelength. The attractive potential generated by the dipole trap laser beams prevents Majorana losses in the node of the magnetic field. The temperature of the trapped atoms is reduced by forced radio-frequency (rf) evaporation. Subsequently, the magnetic field is switched off and the atoms are evaporatively cooled by reducing the laser intensity of the crossed dipole trap. The atoms are being held in the trap for 0.5~s before the ensemble is exposed to a femtosecond laser pulse.

After exposure to the femtosecond laser pulse, the remaining atomic density is mapped onto a CCD camera with a spatial resolution of $(3 \pm 0.5)$~\textmu m by resonant absorption imaging (Fig.~\ref{fig:setup}). An advantage of our setup is the availability of considerably lower temperatures compared to experiments with MOT targets so that we can directly image the ionization vacancy imprinted into the atomic density in real space. During the exposure time of the 50~\textmu s long imaging light pulse the motion of atoms at nanokelvin temperatures is negligible because a typical distance covered is on the order of a few hundred nanometers which is far below the imaging resolution. The detection is performed at the $D_2$ transition of $^{87}$Rb (780~nm wavelength). In the data processing, an exposure- and saturation correction are applied to account for laser intensity fluctuations and population of excited states during imaging. The optical densities are kept considerably below the maximum detectable optical density before saturation correction OD$_\text{max} = 3.7$ given by the dynamic range and the noise level of the 12-bit CCD camera.

The parameters of both ultracold ensembles for non-resonant and resonant ionization are summarized with the imaging parameters in Table~\ref{tab:ultracold_atoms}.

\begin{table}
 \caption{\label{tab:ultracold_atoms} Parameters of the ultracold atomic $^{87}$Rb ensembles for non-resonant ionization at 511~nm and resonant ionization at 1022~nm.}
 \begin{ruledtabular}
  \begin{tabular}{ r r@{$\;\;\pm$}l r@{$\;\;\pm$}l }
                                            & \multicolumn{2}{c}{\textbf{511~nm}} & \multicolumn{2}{c}{\textbf{1022~nm}} \\
	 Temperature $T$ (nK)                     & 109    & 5          &  90    & 8\\
	 Atoms $N_0$ ($10^4$)                     &   1.43 & 0.05       &   1.53 & 0.11\\
	 \hline
   Axial trap frequency $\nu_a$ (Hz)        &  26    & 3          &  27    & 2\\
   Radial trap frequency $\nu_r$ (Hz)       & 129    & 9          & 211    & 2\\
   \hline
   Axial could size $\sigma_a$ (\textmu m)  & 23.9   & 0.4        &  26.3  & 0.9\\
	 Radial could size $\sigma_r$ (\textmu m) &  6.94  & 0.09       &  10.1  & 0.2\\
	 \hline
	 Det. intensity $I_\text{det}$ (mW/cm$^2$)&  1.8   & 0.3        & 2.3 & 0.3\\
	 Max. optical density OD$_\text{max}$     &  \multicolumn{2}{c}{3.0}            & \multicolumn{2}{c}{1.2}
  \end{tabular}
 \end{ruledtabular}
\end{table}

\subsection*{Femtosecond laser pulses}
\label{mth:fs}

The femtosecond laser pulses are generated by a mode-locked commercial chirped-pulse amplification (CPA) ytterbium doped potassium-gadolinium tungstate (Yb:KGW) solid-state laser system running at 1022~nm central wavelength and $\Delta \lambda = 4.36$~nm FWHM bandwidth. This corresponds to a photon energy of $E_\text{ph} = h c / \lambda = 1.21$~eV and a spectral bandwidth of $\Delta E_\text{ph} = (h c / \lambda^2) \Delta \lambda = 5.18$~meV FWHM. The energy and bandwidth covered by two or three 1022~nm photons is $2 (E_\text{ph} \pm \Delta E_\text{ph}) = (2.43 \pm 0.01)$~eV and $3 (E_\text{ph} \pm \Delta E_\text{ph}) = (3.64 \pm 0.02)$~eV, respectively.

The laser emits linearly-polarized pulses of $300^{+30}_{-23}$~fs FWHM duration of the Gaussian temporal intensity distribution at a repetition rate of 100~kHz. Isolated laser pulses can be extracted using a pulse-picker driven by a pockels cell in the regenerative amplifier of the laser system. The pulses are focused onto the ultracold atoms by a $f = 200$~mm achromatic lens where the axis of the beam is tilted by $13.7^\circ$ with respect to the camera axis. In case of the four-photon ionization the pulses are directly transported to the focusing lens.

To study ionization yields at 511~nm wavelength the pulses are converted by second-harmonic generation (SHG) in a Beta-Barium-Borate (BBO) crystal resulting in pulses with a central wavelength of 511.4~nm and 1.75~nm FWHM bandwidth. The pulse duration of the Gaussian temporal intensity distribution in this case is $215^{+20}_{-15}$~fs FWHM. Both pulse durations are measured in an intensity autocorrelation.

In case of the non-resonant measurement at 511~nm the pulse is applied in-situ in the optical dipole trap while for the resonant measurement at 1022~nm the pulse is applied after 2~ms time-of-flight (TOF) to avoid ionization losses induced by leakage of high-repetition rate pulses from the oscillator cavity through the pulse picker. Since these pulses are below the conversion threshold for the SHG, leakage is not an issue in the 511~nm measurements.

\begin{figure}
 \includegraphics[width=1\linewidth]{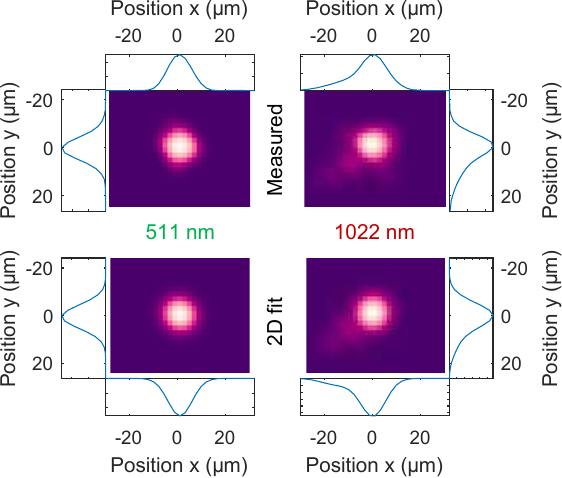}%
 \caption{\label{fig:foci} Focal intensity distribution. Measured (top) and 2D-fitted (bottom) 511~nm (left) and 1022~nm (right) foci with column and row sums on a camera / grid with $2.2 \times 2.2$~\textmu m$^2$ pixel size. The fit model is discussed in the text with the resulting parameters summarized in Table~\ref{tab:focus_fit}.}
\end{figure}

The focus for both wavelengths is characterized using a focal diagnostics camera (compare Fig.~\ref{fig:setup}) with $2.2 \times 2.2$ \textmu m$^2$ pixel size. The pulse energy $E_p$ is recorded simultaneously for each measurement to accurately determine the intensity in the focal volume for calibrated peak intensities $I_0$. Figure~\ref{fig:foci} depicts the focal intensity distribution together with a 2D Gaussian fit for both wavelengths. While the 511~nm focus is very well represented by a single Gaussian intensity distribution, the 1022~nm focus deviates from this ideal model by two additional weak features on the bottom left corner of the image. In order to calculate the loss fraction expected by theory, we reconstruct the intensity distribution using a two-dimensional fit.

The 3D intensity distribution of the laser pulse in space and time in this general case is given by

\begin{multline}
 I(x,y,z,t) = I_0 \exp{\left( - \frac{t^2}{2 \tau^2} \right)}\\
 \times \sum_{k = 1}^{3} A^k \frac{w^k_{0x} w^k_{0y}}{w^k_x(z) w^k_y(z)} \exp{\left(- \left[\frac{2 x^2}{{w^k_x}^2(z)} + \frac{2 \left(y - y_0^k\right)^2}{{w^k_y}^2(z)} \right] \right)}%
 \label{eq:intensity}
\end{multline}

\noindent where the sum accounts for the superposition of up to three elliptical Gaussian profiles $k \in \{ \text{I}, \text{II}, \text{III} \}$ offset along the $y$ axis by $y_0^k$. For each profile $k$ and each coordinate $i \in \{ x, y \}$ the waist is given by $w^k_i(z) = w^k_{0i} \sqrt{1 + (z/z^k_{Ri})^2}$ with the Rayleigh length $z^k_{Ri} = \pi {w^k_{0i}}^2 / \lambda$. The peak intensity 

\begin{equation}
 I_0 = \frac{2 P_0}{\pi \sum_{k = 1}^{3} A^k w^k_{0x} w^k_{0y}}%
 \label{eq:peakintensity}
\end{equation}

\noindent includes the peak power $P_0 = E_p / (\sqrt{2 \pi} \tau)$ with pulse energy $E_p$ and rms pulse duration $\tau = \tau_\text{FWHM} / \left( 2 \sqrt{2 \ln{(2)}} \right)$. The amplitude of the main peak is normalized to $A^\text{I} = 1$ while the amplitudes for the second and third beam ($A^\text{II}$ and $A^\text{III}$) are set to zero in the 511~nm case and determined by the fit in the 1022~nm case. Integration over the spatiotemporal intensity profile returns the pulse energy $E_p$.

The 2D fit is performed using Eq.~(\ref{eq:intensity}) in the focal plane at $z = 0$. A global rotation of the coordinate system around the $z$ axis by an angle $\alpha_z$ is included into the fit. The resulting fit parameters with confidence intervals are summarized in Table~\ref{tab:focus_fit} and the reconstruction is plotted in the bottom right display of Fig.~\ref{fig:foci}.

\begin{table}
 \caption{\label{tab:focus_fit} Two-dimensional fit parameters with confidence intervals of the measured focal intensity distributions for 511~nm and 1022~nm wavelength as presented in Fig.~\ref{fig:foci}.}
 \begin{ruledtabular}
  \begin{tabular}{ r r@{$\;\pm$}l r@{$\;\pm$}l }
                          & \multicolumn{2}{c}{\textbf{511~nm}} & \multicolumn{2}{c}{\hspace{1.5cm}\textbf{1022~nm}} \\
   $w^\text{I}_{0x}$ (\textmu m) \hspace{1.5cm} & 10.22 & 0.07 & \hspace{1.5cm} 10.3 & 0.1\\
   $w^\text{I}_{0y}$ (\textmu m) \hspace{1.5cm} &  9.67 & 0.07 &                10.5 & 0.1\\
   \hline
   $A^\text{II}$                  \hspace{1.5cm} & \multicolumn{2}{c}{0}    &   0.14 & 0.04\\
   $w^\text{II}_{0x}$ (\textmu m) \hspace{1.5cm} & \multicolumn{2}{c}{-/-}  &  15    & 2\\
   $w^\text{II}_{0y}$ (\textmu m) \hspace{1.5cm} & \multicolumn{2}{c}{-/-}  &   5.0  & 0.9\\
	 $y^\text{II}_0$    (\textmu m) \hspace{1.5cm} & \multicolumn{2}{c}{-/-}  & -13.5  & 0.4\\
   \hline
   $A^\text{III}$                  \hspace{1.5cm} & \multicolumn{2}{c}{0}   &   0.12 & 0.01\\
   $w^\text{III}_{0x}$ (\textmu m) \hspace{1.5cm} & \multicolumn{2}{c}{-/-} &   9    & 1\\
   $w^\text{III}_{0y}$ (\textmu m) \hspace{1.5cm} & \multicolumn{2}{c}{-/-} &  11    & 3\\
	 $y^\text{III}_0$    (\textmu m) \hspace{1.5cm} & \multicolumn{2}{c}{-/-} & -22    & 2\\
   \hline
   $\alpha_z (^\circ)$             \hspace{1.5cm} & 37 & 5                  & -126   & 1
  \end{tabular}
 \end{ruledtabular}
\end{table}

\subsection*{Time-dependent Schr\"odinger equation}
\label{mth:tdse}

Ionization of rubidium is investigated theoretically by numerically solving the Time Dependent Schr\"odinger Equation (TDSE) within the single active electron approximation. In brief, TDSE is treated in spherical coordinates within the split-propagation paradigm. Angular dynamics is described within the basis of 20 spherical $\sigma$ harmonics. The propagation step over the non-uniform grid of the radial coordinate is performed with the Crank-Nicolson algorithm.

The $l$-dependent interaction of the electron with the residual positive rubidium ion is parametrized to reproduce energies of the $5s$ ground state and six lowest excited states of each $s, p, d$, and $f$ series. The average over the corresponding NIST tabulated multiplet energies is taken as experimental energies of the rubidium states. The eigenenergies of the used potentials reproduce the experimental values with $10^{-4}$ accuracy.

The region of the wave packet propagation is 5200~a.u. Initially, only the $5s$ state is populated and the intensity of the Gauss shaped excitation pulse of 215~fs FWHM duration for 511~nm and 300~fs for 1022~nm is very small at this moment. An absorbing potential at the boundary of the radial mesh is used for suppression of wave packet reflection from this boundary. The propagation time is substantially larger than the pulse duration and therefore a quite noticeable decrease of the wave packet population at $r \le 400$~a.u. is obtained.

The results of the propagation are the time-dependent populations of the involved excited states. As the full probability of the system survival after ionization, we took the full wave packet population inside the sphere $r \le 400$~a.u. The part of the wave packet beyond this sphere is very small and contains contributions from slowly receding electrons and high Rydberg states. An analysis of the results reveals the role of resonant states in the case of a laser pulse of 1022~nm wavelength. Applying a laser pulse of 511~nm wavelength does not excite resonances.

\subsection*{Calculation of the loss fraction}
\label{mth:calculation}

Using ultracold atoms, a local 2D density distribution of the atoms is accessible via the absorption imaging technique. In order to calculate the fraction of lost atoms and compare it to different ionization models, a double Gaussian function according to Eq.~(\ref{eq:doublegaussian}) is fitted to the row sum of the absorption image where $A_0$ determines an offset while $A_c$, $x_c$ and $\sigma_c$ correspond to amplitude, central position and width of the atomic cloud and $A_v$, $x_v$ and $\sigma_v$ describe the same parameters for the vacancy, respectively. The fraction of lost atoms can be derived by Eq.~(\ref{eq:fl}). For an example the reader is referred to Fig.~\ref{fig:clouds}.

In order to calculate the expected loss fraction, the full 3D initial atomic density

\begin{equation}
 \rho_i(x,y,z) = \frac{N_0}{(2 \pi)^\frac{3}{2} \sigma_a \sigma_r^2} \exp{\left(- \frac{x^2}{2 \sigma_a^2} \right)} \exp{\left(- \frac{y^2 + z^2}{2 \sigma_r^2} \right)}%
 \label{eq:rhoi}
\end{equation}

\noindent is modeled by a Gaussian cloud consisting of $N_0$ atoms distributed within $\sigma_a$ axial rms width and $\sigma_r$ radial rms width, all given in Table~\ref{tab:ultracold_atoms}.

From the 3D intensity distribution in Eq.~(\ref{eq:intensity}) the ionization probability $P(x,y,z)$ after the laser pulse is obtained for each grid unit. In case of the TDSE results the ionization probability presented in Fig.~\ref{fig:data_511nm}a and Fig.~\ref{fig:data_1022nm}a is used. For the MPI, OBI and ADK models the ionization probability is calculated as demonstrated in the Supplemental Material. This ionization probability map is now superimposed with the initial 3D atomic density map in Eq.~(\ref{eq:rhoi}) and $N_i = P N_0$ atoms at a specific grid unit are removed from the initial 3D density leading to the final density

\begin{equation}
 \rho_f(x,y,z) = \rho_i(x,y,z) \left[ 1 - P(x,y,z) \right]%
 \label{eq:rhof}
\end{equation}

\noindent after ionization.

Subsequently, this 3D density distribution $\rho_f(x',y',z')$ is projected along the imaging axis in order to obtain a simulated 2D density similar to the absorption image. The small angle between the ionizing laser beam and the camera axis of $13.7^\circ$ has been taken into account. Therefore, the 2D coordinate system is rotated by this angle with respect to the 3D coordinate system. The 2D density distribution $\rho_\text{2D}(x,y)$ is convoluted with the optical resolution of the imaging system and evaluated in the same manner as the absorption image. The row sum $\rho_\text{1D}(x)$ is calculated and fitted to Eq.~(\ref{eq:doublegaussian}) which results in a simulated loss fraction $f_l^\text{sim}$ extracted from the fit according to Eq.~(\ref{eq:fl}).

A simulated absorption image compared to the corresponding experimental image and the projected row sum is presented in Fig.~\ref{fig:clouds}. The result for the simulated loss fractions with respect to the intensity is given in Fig.~\ref{fig:data_511nm}b and Fig.~\ref{fig:data_1022nm}b. All calculations were performed on a three dimensional grid ($x \times y \times z$) of $500 \times 120 \times 120$ grid units each $1 \times 1 \times 1$ \textmu m$^3$ in size so that the total simulated area spans $500 \times 120 \times 120$ \textmu m$^3$.

\bibliography{rb_strongfield}

\begin{thebibliography}{10}
\expandafter\ifx\csname url\endcsname\relax
  \def\url#1{\texttt{#1}}\fi
\expandafter\ifx\csname urlprefix\endcsname\relax\def\urlprefix{URL }\fi
\providecommand{\bibinfo}[2]{#2}
\providecommand{\eprint}[2][]{\url{#2}}

\bibitem{Tomza:ColdAtomIonReview}
\bibinfo{author}{Tomza, M.} \emph{et~al.}
\newblock \bibinfo{title}{Cold hybrid ion-atom systems} (\bibinfo{year}{2017}).
\newblock \urlprefix\url{https://arxiv.org/abs/1708.07832}.
\newblock \eprint{arXiv:1708.07832}.

\bibitem{Haerter:ColdAtomIonReview}
\bibinfo{author}{H\"arter, A.} \& \bibinfo{author}{Denschlag, J.~H.}
\newblock \bibinfo{title}{Cold atom-ion experiments in hybrid traps}.
\newblock \emph{\bibinfo{journal}{Contemp. Phys.}}
  \textbf{\bibinfo{volume}{55}}, \bibinfo{pages}{33--45}
  (\bibinfo{year}{2014}).
\newblock \urlprefix\url{https://doi.org/10.1080/00107514.2013.854618}.

\bibitem{Doerk:AtomIonQG}
\bibinfo{author}{Doerk, H.}, \bibinfo{author}{Idziaszek, Z.} \&
  \bibinfo{author}{Calarco, T.}
\newblock \bibinfo{title}{Atom-ion quantum gate}.
\newblock \emph{\bibinfo{journal}{Phys. Rev. A}} \textbf{\bibinfo{volume}{81}},
  \bibinfo{pages}{012708} (\bibinfo{year}{2010}).
\newblock \urlprefix\url{https://doi.org/10.1103/PhysRevA.81.012708}.

\bibitem{Idziaszek:AtomIonCollisions}
\bibinfo{author}{Idziaszek, Z.}, \bibinfo{author}{Calarco, T.} \&
  \bibinfo{author}{Zoller, P.}
\newblock \bibinfo{title}{Controlled collisions of a single atom and an ion
  guided by movable trapping potentials}.
\newblock \emph{\bibinfo{journal}{Phys. Rev. A}} \textbf{\bibinfo{volume}{76}},
  \bibinfo{pages}{033409} (\bibinfo{year}{2007}).
\newblock \urlprefix\url{https://doi.org/10.1103/PhysRevA.76.033409}.

\bibitem{Schurer:MesoscopicIon}
\bibinfo{author}{Schurer, J.~M.}, \bibinfo{author}{Negretti, A.} \&
  \bibinfo{author}{Schmelcher, P.}
\newblock \bibinfo{title}{Unraveling the structure of ultracold mesoscopic
  collinear molecular ions}.
\newblock \emph{\bibinfo{journal}{Phys. Rev. Lett.}}
  \textbf{\bibinfo{volume}{119}}, \bibinfo{pages}{063001}
  (\bibinfo{year}{2017}).
\newblock \urlprefix\url{https://doi.org/10.1103/PhysRevLett.119.063001}.

\bibitem{Schurer:IonCapture}
\bibinfo{author}{Schurer, J.~M.}, \bibinfo{author}{Negretti, A.} \&
  \bibinfo{author}{Schmelcher, P.}
\newblock \bibinfo{title}{Capture dynamics of ultracold atoms in the presence
  of an impurity ion}.
\newblock \emph{\bibinfo{journal}{New J. Phys.}} \textbf{\bibinfo{volume}{17}},
  \bibinfo{pages}{083024} (\bibinfo{year}{2015}).
\newblock \urlprefix\url{https://doi.org/10.1088/1367-2630/17/8/083024}.

\bibitem{Killian:Plasma}
\bibinfo{author}{Killian, T.~C.} \emph{et~al.}
\newblock \bibinfo{title}{Creation of an ultracold neutral plasma}.
\newblock \emph{\bibinfo{journal}{Phys. Rev. Lett.}}
  \textbf{\bibinfo{volume}{83}}, \bibinfo{pages}{4776--4779}
  (\bibinfo{year}{1999}).
\newblock \urlprefix\url{https://doi.org/10.1103/PhysRevLett.83.4776}.

\bibitem{Killian:PlasmaReview}
\bibinfo{author}{Killian, T.~C.}
\newblock \bibinfo{title}{Ultracold neutral plasmas}.
\newblock \emph{\bibinfo{journal}{Science}} \textbf{\bibinfo{volume}{316}},
  \bibinfo{pages}{705--708} (\bibinfo{year}{2007}).
\newblock \urlprefix\url{https://doi.org/10.1126/science.1130556}.

\bibitem{Killian:UNPPhysReports}
\bibinfo{author}{Killian, T.}, \bibinfo{author}{Pattard, T.},
  \bibinfo{author}{Pohl, T.} \& \bibinfo{author}{Rost, J.}
\newblock \bibinfo{title}{Ultracold neutral plasmas}.
\newblock \emph{\bibinfo{journal}{Phys. Rep.}} \textbf{\bibinfo{volume}{449}},
  \bibinfo{pages}{77 -- 130} (\bibinfo{year}{2007}).
\newblock \urlprefix\url{https://doi.org/10.1016/j.physrep.2007.04.007}.

\bibitem{Murphy:RbIonsSC}
\bibinfo{author}{Murphy, D.} \emph{et~al.}
\newblock \bibinfo{title}{Detailed observation of space-charge dynamics using
  ultracold ion bunches}.
\newblock \emph{\bibinfo{journal}{Nat. Commun.}} \textbf{\bibinfo{volume}{5}},
  \bibinfo{pages}{--} (\bibinfo{year}{2014}).
\newblock \urlprefix\url{https://doi.org/10.1038/ncomms5489}.

\bibitem{Takei:ManyBodyElDynamics}
\bibinfo{author}{Takei, N.} \emph{et~al.}
\newblock \bibinfo{title}{Direct observation of ultrafast many-body electron
  dynamics in an ultracold rydberg gas}.
\newblock \emph{\bibinfo{journal}{Nat. Commun.}} \textbf{\bibinfo{volume}{7}},
  \bibinfo{pages}{13449} (\bibinfo{year}{2016}).
\newblock \urlprefix\url{https://doi.org/10.1038/ncomms13449}.

\bibitem{Claessens:UltracoldElectronSource}
\bibinfo{author}{Claessens, B.~J.}, \bibinfo{author}{van~der Geer, S.~B.},
  \bibinfo{author}{Taban, G.}, \bibinfo{author}{Vredenbregt, E. J.~D.} \&
  \bibinfo{author}{Luiten, O.~J.}
\newblock \bibinfo{title}{Ultracold electron source}.
\newblock \emph{\bibinfo{journal}{Phys. Rev. Lett.}}
  \textbf{\bibinfo{volume}{95}}, \bibinfo{pages}{164801}
  (\bibinfo{year}{2005}).
\newblock \urlprefix\url{https://doi.org/10.1103/PhysRevLett.95.164801}.

\bibitem{McCulloch:ElectronBunches}
\bibinfo{author}{McCulloch, A.~J.}, \bibinfo{author}{Sheludko, D.~V.},
  \bibinfo{author}{Junker, M.} \& \bibinfo{author}{Scholten, R.~E.}
\newblock \bibinfo{title}{High-coherence picosecond electron bunches from cold
  atoms}.
\newblock \emph{\bibinfo{journal}{Nat. Commun.}} \textbf{\bibinfo{volume}{4}},
  \bibinfo{pages}{1692} (\bibinfo{year}{2013}).
\newblock \urlprefix\url{https://doi.org/10.1038/ncomms2699}.

\bibitem{Protopapas:HighIntensityReview}
\bibinfo{author}{Protopapas, M.}, \bibinfo{author}{Keitel, C.~H.} \&
  \bibinfo{author}{Knight, P.~L.}
\newblock \bibinfo{title}{Atomic physics with super-high intensity lasers}.
\newblock \emph{\bibinfo{journal}{Rep. Prog. Phys.}}
  \textbf{\bibinfo{volume}{60}}, \bibinfo{pages}{389} (\bibinfo{year}{1997}).
\newblock \urlprefix\url{https://doi.org/10.1088/0034-4885/60/4/001}.

\bibitem{Agostini:IonizationStrongFieldReview}
\bibinfo{author}{Agostini, P.} \& \bibinfo{author}{DiMauro, L.~F.}
\newblock \bibinfo{title}{Atomic and molecular ionization dynamics in strong
  laser fields: From optical to x-rays}.
\newblock \emph{\bibinfo{journal}{Adv. At., Mol., Opt. Phys.}}
  \textbf{\bibinfo{volume}{61}}, \bibinfo{pages}{117 -- 158}
  (\bibinfo{year}{2012}).
\newblock \urlprefix\url{https://doi.org/10.1016/B978-0-12-396482-3.00003-X}.

\bibitem{Keldysh:Ionization}
\bibinfo{author}{Keldysh, L.}
\newblock \bibinfo{title}{Ionization in the field of a strong electromagnetic
  wave}.
\newblock \emph{\bibinfo{journal}{Sov. Phys. JETP}}
  \textbf{\bibinfo{volume}{20}}, \bibinfo{pages}{1307} (\bibinfo{year}{1965}).
\newblock
  \urlprefix\url{http://jetp.ac.ru/cgi-bin/e/index/e/20/5/p1307?a=list}.

\bibitem{PPT:Ionization}
\bibinfo{author}{Perelomov, A.~M.}, \bibinfo{author}{Popov, V.~S.} \&
  \bibinfo{author}{Terent'ev, M.~V.}
\newblock \bibinfo{title}{Ionization of atoms in an alternating electric
  field}.
\newblock \emph{\bibinfo{journal}{Sov. Phys. JETP}}
  \textbf{\bibinfo{volume}{23}}, \bibinfo{pages}{924} (\bibinfo{year}{1966}).
\newblock
  \urlprefix\url{http://www.jetp.ac.ru/cgi-bin/e/index/e/23/5/p924?a=list}.

\bibitem{ADK:Tunnel}
\bibinfo{author}{Ammosov, M.}, \bibinfo{author}{Delone, N.} \&
  \bibinfo{author}{Krainov, V.}
\newblock \bibinfo{title}{Tunnel ionization of complex atoms and of atomic ions
  in an alternating electromagnetic field}.
\newblock \emph{\bibinfo{journal}{Sov. Phys. JETP}}
  \textbf{\bibinfo{volume}{64}}, \bibinfo{pages}{1191} (\bibinfo{year}{1986}).
\newblock
  \urlprefix\url{http://jetp.ac.ru/cgi-bin/e/index/e/64/6/p1191?a=list}.

\bibitem{Augst:TunnelIonization}
\bibinfo{author}{Augst, S.}, \bibinfo{author}{Strickland, D.},
  \bibinfo{author}{Meyerhofer, D.~D.}, \bibinfo{author}{Chin, S.~L.} \&
  \bibinfo{author}{Eberly, J.~H.}
\newblock \bibinfo{title}{Tunneling ionization of noble gases in a
  high-intensity laser field}.
\newblock \emph{\bibinfo{journal}{Phys. Rev. Lett.}}
  \textbf{\bibinfo{volume}{63}}, \bibinfo{pages}{2212--2215}
  (\bibinfo{year}{1989}).
\newblock \urlprefix\url{https://doi.org/10.1103/PhysRevLett.63.2212}.

\bibitem{Schuricke:StrongFieldLi}
\bibinfo{author}{Schuricke, M.} \emph{et~al.}
\newblock \bibinfo{title}{Strong-field ionization of lithium}.
\newblock \emph{\bibinfo{journal}{Phys. Rev. A}} \textbf{\bibinfo{volume}{83}},
  \bibinfo{pages}{023413} (\bibinfo{year}{2011}).
\newblock \urlprefix\url{https://doi.org/10.1103/PhysRevA.83.023413}.

\bibitem{Schuricke:LiCoherence}
\bibinfo{author}{Schuricke, M.} \emph{et~al.}
\newblock \bibinfo{title}{Coherence in multistate resonance-enhanced
  four-photon ionization of lithium atoms}.
\newblock \emph{\bibinfo{journal}{Phys. Rev. A}} \textbf{\bibinfo{volume}{88}},
  \bibinfo{pages}{023427} (\bibinfo{year}{2013}).
\newblock \urlprefix\url{https://doi.org/10.1103/PhysRevA.88.023427}.

\bibitem{Hart:SFIBSINa}
\bibinfo{author}{Hart, N.~A.} \emph{et~al.}
\newblock \bibinfo{title}{Selective strong-field enhancement and suppression of
  ionization with short laser pulses}.
\newblock \emph{\bibinfo{journal}{Phys. Rev. A}} \textbf{\bibinfo{volume}{93}},
  \bibinfo{pages}{063426} (\bibinfo{year}{2016}).
\newblock \urlprefix\url{https://doi.org/10.1103/PhysRevA.93.063426}.

\bibitem{Boge:AttoclockSFI}
\bibinfo{author}{Boge, R.} \emph{et~al.}
\newblock \bibinfo{title}{Probing nonadiabatic effects in strong-field tunnel
  ionization}.
\newblock \emph{\bibinfo{journal}{Phys. Rev. Lett.}}
  \textbf{\bibinfo{volume}{111}}, \bibinfo{pages}{103003}
  (\bibinfo{year}{2013}).
\newblock \urlprefix\url{https://doi.org/10.1103/PhysRevLett.111.103003}.

\bibitem{Hofmann:NonAdiabaticSF}
\bibinfo{author}{Hofmann, C.}, \bibinfo{author}{Zimmermann, T.},
  \bibinfo{author}{Zielinski, A.} \& \bibinfo{author}{Landsman, A.~S.}
\newblock \bibinfo{title}{Non-adiabatic imprints on the electron wave packet in
  strong field ionization with circular polarization}.
\newblock \emph{\bibinfo{journal}{New J. Phys.}} \textbf{\bibinfo{volume}{18}},
  \bibinfo{pages}{043011} (\bibinfo{year}{2016}).
\newblock \urlprefix\url{https://doi.org/10.1088/1367-2630/18/4/043011}.

\bibitem{Dinneen:TrapLoss}
\bibinfo{author}{Dinneen, T.~P.}, \bibinfo{author}{Wallace, C.~D.},
  \bibinfo{author}{Tan, K.-Y.~N.} \& \bibinfo{author}{Gould, P.~L.}
\newblock \bibinfo{title}{Use of trapped atoms to measure absolute
  photoionization cross sections}.
\newblock \emph{\bibinfo{journal}{Opt. Lett.}} \textbf{\bibinfo{volume}{17}},
  \bibinfo{pages}{1706--1708} (\bibinfo{year}{1992}).
\newblock \urlprefix\url{https://doi.org/10.1364/OL.17.001706}.

\bibitem{Madsen:AbsoluteMgCS}
\bibinfo{author}{Madsen, D.~N.} \& \bibinfo{author}{Thomsen, J.~W.}
\newblock \bibinfo{title}{Measurement of absolute photo-ionization cross
  sections using magnesium magneto-optical traps}.
\newblock \emph{\bibinfo{journal}{J. Phys. B: At., Mol. Opt. Phys.}}
  \textbf{\bibinfo{volume}{35}}, \bibinfo{pages}{2173} (\bibinfo{year}{2002}).
\newblock \urlprefix\url{https://doi.org/10.1088/0953-4075/35/9/314}.

\bibitem{Anderlini:TwoPhotonColdRb}
\bibinfo{author}{Anderlini, M.} \emph{et~al.}
\newblock \bibinfo{title}{Two-photon ionization of cold rubidium atoms}.
\newblock \emph{\bibinfo{journal}{J. Opt. Soc. Am. B}}
  \textbf{\bibinfo{volume}{21}}, \bibinfo{pages}{480--485}
  (\bibinfo{year}{2004}).
\newblock \urlprefix\url{https://doi.org/10.1364/JOSAB.21.000480}.

\bibitem{Lin:HybridTrap}
\bibinfo{author}{Lin, Y.-J.}, \bibinfo{author}{Perry, A.~R.},
  \bibinfo{author}{Compton, R.~L.}, \bibinfo{author}{Spielman, I.~B.} \&
  \bibinfo{author}{Porto, J.~V.}
\newblock \bibinfo{title}{Rapid production of $^{87}\text{R}\text{b}$
  bose-einstein condensates in a combined magnetic and optical potential}.
\newblock \emph{\bibinfo{journal}{Phys. Rev. A}} \textbf{\bibinfo{volume}{79}},
  \bibinfo{pages}{063631} (\bibinfo{year}{2009}).
\newblock \urlprefix\url{https://doi.org/10.1103/PhysRevA.79.063631}.

\bibitem{Takekoshi:RbMPI}
\bibinfo{author}{Takekoshi, T.}, \bibinfo{author}{Brooke, G.~M.},
  \bibinfo{author}{Patterson, B.~M.} \& \bibinfo{author}{Knize, R.~J.}
\newblock \bibinfo{title}{Absolute rb one-color two-photon ionization
  cross-section measurement near a quantum interference}.
\newblock \emph{\bibinfo{journal}{Phys. Rev. A}} \textbf{\bibinfo{volume}{69}},
  \bibinfo{pages}{053411} (\bibinfo{year}{2004}).
\newblock \urlprefix\url{https://doi.org/10.1103/PhysRevA.69.053411}.

\bibitem{Morishita:LiSFTheory}
\bibinfo{author}{Morishita, T.} \& \bibinfo{author}{Lin, C.~D.}
\newblock \bibinfo{title}{Photoelectron spectra and high rydberg states of
  lithium generated by intense lasers in the over-the-barrier ionization
  regime}.
\newblock \emph{\bibinfo{journal}{Phys. Rev. A}} \textbf{\bibinfo{volume}{87}},
  \bibinfo{pages}{063405} (\bibinfo{year}{2013}).
\newblock \urlprefix\url{https://doi.org/10.1103/PhysRevA.87.063405}.

\bibitem{Jheng:LiIonTheory}
\bibinfo{author}{Jheng, S.-D.} \& \bibinfo{author}{Jiang, T.~F.}
\newblock \bibinfo{title}{A theoretical study on the strong-field ionization of
  the lithium atom}.
\newblock \emph{\bibinfo{journal}{J. Phys. B: At., Mol. Opt. Phys.}}
  \textbf{\bibinfo{volume}{46}}, \bibinfo{pages}{115601}
  (\bibinfo{year}{2013}).
\newblock \urlprefix\url{https://doi.org/10.1088/0953-4075/46/11/115601}.

\bibitem{DK:TunnelBSI}
\bibinfo{author}{Delone, N.~B.} \& \bibinfo{author}{Krainov, V.~P.}
\newblock \bibinfo{title}{Tunneling and barrier-suppression ionization of atoms
  and ions in a laser radiation field}.
\newblock \emph{\bibinfo{journal}{Phys. Usp.}} \textbf{\bibinfo{volume}{41}},
  \bibinfo{pages}{469} (\bibinfo{year}{1998}).
\newblock \urlprefix\url{https://doi.org/10.1070/PU1998v041n05ABEH000393}.

\bibitem{Safronova:Rubidium}
\bibinfo{author}{Safronova, M.~S.} \& \bibinfo{author}{Safronova, U.~I.}
\newblock \bibinfo{title}{Critically evaluated theoretical energies, lifetimes,
  hyperfine constants, and multipole polarizabilities in $^{87}\mathrm{Rb}$}.
\newblock \emph{\bibinfo{journal}{Phys. Rev. A}} \textbf{\bibinfo{volume}{83}},
  \bibinfo{pages}{052508} (\bibinfo{year}{2011}).
\newblock \urlprefix\url{https://doi.org/10.1103/PhysRevA.83.052508}.

\bibitem{Jheng:LiWavePacketDisplacement}
\bibinfo{author}{Jheng, S.-D.} \& \bibinfo{author}{Jiang, T.~F.}
\newblock \bibinfo{title}{Effect of bound electron wave packet displacement on
  the multiphoton ionization of a lithium atom}.
\newblock \emph{\bibinfo{journal}{J. Phys. B: At., Mol. Opt. Phys.}}
  \textbf{\bibinfo{volume}{50}}, \bibinfo{pages}{195001}
  (\bibinfo{year}{2017}).
\newblock \urlprefix\url{https://doi.org/10.1088/1361-6455/aa8968}.

\bibitem{Sharma:AllOpticalMOT}
\bibinfo{author}{Sharma, S.} \emph{et~al.}
\newblock \bibinfo{title}{An all-optical atom trap as a target for motrims-like
  collision experiments} (\bibinfo{year}{2017}).
\newblock \urlprefix\url{https://arxiv.org/abs/1712.01200}.
\newblock \eprint{arXiv:1712.01200}.

\bibitem{Henkel:SingleIonDetection}
\bibinfo{author}{Henkel, F.} \emph{et~al.}
\newblock \bibinfo{title}{Highly efficient state-selective submicrosecond
  photoionization detection of single atoms}.
\newblock \emph{\bibinfo{journal}{Phys. Rev. Lett.}}
  \textbf{\bibinfo{volume}{105}}, \bibinfo{pages}{253001}
  (\bibinfo{year}{2010}).
\newblock \urlprefix\url{https://doi.org/10.1103/PhysRevLett.105.253001}.

\bibitem{Stecker:ColdIonMicroscope}
\bibinfo{author}{Stecker, M.}, \bibinfo{author}{Schefzyk, H.},
  \bibinfo{author}{Fort\'agh, J.} \& \bibinfo{author}{G\"unther, A.}
\newblock \bibinfo{title}{A high resolution ion microscope for cold atoms}.
\newblock \emph{\bibinfo{journal}{New Journal of Physics}}
  \textbf{\bibinfo{volume}{19}}, \bibinfo{pages}{043020}
  (\bibinfo{year}{2017}).
\newblock \urlprefix\url{https://doi.org/10.1088/1367-2630/aa6741}.

\bibitem{Goetz:UltracoldIonizationLED}
\bibinfo{author}{G\"otz, S.}, \bibinfo{author}{H\"oltkemeier, B.},
  \bibinfo{author}{Amthor, T.} \& \bibinfo{author}{Weidem\"uller, M.}
\newblock \bibinfo{title}{Photoionization of optically trapped ultracold atoms
  with a high-power light-emitting diode}.
\newblock \emph{\bibinfo{journal}{Rev. Sci. Instrum.}}
  \textbf{\bibinfo{volume}{84}}, \bibinfo{pages}{043107}
  (\bibinfo{year}{2013}).
\newblock \urlprefix\url{https://doi.org/10.1063/1.4795475}.

\bibitem{Milosevic:StrongFieldAlkali}
\bibinfo{author}{Milo\ifmmode \check{s}\else
  \v{s}\fi{}evi\ifmmode~\acute{c}\else \'{c}\fi{}, M.~Z.} \&
  \bibinfo{author}{Simonovi\ifmmode~\acute{c}\else \'{c}\fi{}, N.~S.}
\newblock \bibinfo{title}{Calculations of rates for strong-field ionization of
  alkali-metal atoms in the quasistatic regime}.
\newblock \emph{\bibinfo{journal}{Phys. Rev. A}} \textbf{\bibinfo{volume}{91}},
  \bibinfo{pages}{023424} (\bibinfo{year}{2015}).
\newblock \urlprefix\url{https://doi.org/10.1103/PhysRevA.91.023424}.

\bibitem{Bunjac:SFQCSodium}
\bibinfo{author}{Bunjac, A.}, \bibinfo{author}{Popovi\'{c}, D.~B.} \&
  \bibinfo{author}{Simonovi\'{c}, N.~S.}
\newblock \bibinfo{title}{Resonant dynamic stark shift as a tool in
  strong-field quantum control: calculation and application for selective
  multiphoton ionization of sodium}.
\newblock \emph{\bibinfo{journal}{Phys. Chem. Chem. Phys.}}
  \textbf{\bibinfo{volume}{19}}, \bibinfo{pages}{19829--19836}
  (\bibinfo{year}{2017}).
\newblock \urlprefix\url{https://doi.org/10.1039/C7CP02146A}.

\bibitem{Miladinovic:AlkaliBSI}
\bibinfo{author}{Miladinovi{\'{c}}, T.~B.} \& \bibinfo{author}{Petrovi{\'{c}},
  V.~M.}
\newblock \bibinfo{title}{Laser field ionization rates in the
  barrier-suppression regime}.
\newblock \emph{\bibinfo{journal}{J. Russ. Laser Res.}}
  \textbf{\bibinfo{volume}{36}}, \bibinfo{pages}{312--319}
  (\bibinfo{year}{2015}).
\newblock \urlprefix\url{https://doi.org/10.1007/s10946-015-9505-0}.

\bibitem{Krug:CoherentStrongFieldControl}
\bibinfo{author}{Krug, M.} \emph{et~al.}
\newblock \bibinfo{title}{Coherent strong-field control of multiple states by a
  single chirped femtosecond laser pulse}.
\newblock \emph{\bibinfo{journal}{New J. Phys.}} \textbf{\bibinfo{volume}{11}},
  \bibinfo{pages}{105051} (\bibinfo{year}{2009}).
\newblock \urlprefix\url{https://doi.org/10.1088/1367-2630/11/10/105051}.

\bibitem{Ciampini:BECPhotoionization}
\bibinfo{author}{Ciampini, D.} \emph{et~al.}
\newblock \bibinfo{title}{Photoionization of ultracold and
  bose-einstein-condensed rb atoms}.
\newblock \emph{\bibinfo{journal}{Phys. Rev. A}} \textbf{\bibinfo{volume}{66}},
  \bibinfo{pages}{043409} (\bibinfo{year}{2002}).
\newblock \urlprefix\url{https://doi.org/10.1103/PhysRevA.66.043409}.

\bibitem{Mazets:BECIonizationSuppression}
\bibinfo{author}{Mazets, I.~E.}
\newblock \bibinfo{title}{Photoionization of neutral atoms in a bose-einstein
  condensate}.
\newblock \emph{\bibinfo{journal}{Quantum Semiclass. Opt.}}
  \textbf{\bibinfo{volume}{10}}, \bibinfo{pages}{675} (\bibinfo{year}{1998}).
\newblock \urlprefix\url{https://doi.org/10.1088/1355-5111/10/5/005}.

\bibitem{Edwards:MPI5Rb}
\bibinfo{author}{Edwards, M.}, \bibinfo{author}{Tang, X.} \&
  \bibinfo{author}{Shakeshaft, R.}
\newblock \bibinfo{title}{Multiphoton absorption by alkali-metal atoms above
  the ionization threshold. ii. further results on cs and rb ionization}.
\newblock \emph{\bibinfo{journal}{Phys. Rev. A}} \textbf{\bibinfo{volume}{35}},
  \bibinfo{pages}{3758--3767} (\bibinfo{year}{1987}).
\newblock \urlprefix\url{https://doi.org/10.1103/PhysRevA.35.3758}.

\bibitem{Delone:Rb4PhotonCS}
\bibinfo{author}{Delone, G.~A.}, \bibinfo{author}{Manakov, N.~L.},
  \bibinfo{author}{Preobrazhenskii, M.~A.} \& \bibinfo{author}{Rapoport, L.}
\newblock \bibinfo{title}{Polarization effects in multiphoton ionization of
  alkali atoms}.
\newblock \emph{\bibinfo{journal}{Sov. Phys. JETP}}
  \textbf{\bibinfo{volume}{43}}, \bibinfo{pages}{642} (\bibinfo{year}{1976}).
\newblock \urlprefix\url{http://jetp.ac.ru/cgi-bin/e/index/e/43/4/p642?a=list}.

\bibitem{Spielmann:keVXray}
\bibinfo{author}{Spielmann, C.} \emph{et~al.}
\newblock \bibinfo{title}{Near-kev coherent x-ray generation with sub-10-fs
  lasers}.
\newblock \emph{\bibinfo{journal}{IEEE J. Sel. Topics in Quantum Electron.}}
  \textbf{\bibinfo{volume}{4}}, \bibinfo{pages}{249--265}
  (\bibinfo{year}{1998}).
\newblock \urlprefix\url{https://doi.org/10.1109/2944.686730}.

\end{thebibliography}

\begin{acknowledgments}
 \label{sec:acknowledgments}
 This work has been supported by the Cluster of Excellence 'The Hamburg Centre for Ultrafast Imaging (CUI)' of the German Science Foundation (DFG). We thank Harald Blazy, Jakob Butlewski, Hannes Duncker, Alexander Grote, Jasper Krauser and Harry Kr\"uger for contributions during an early stage of the experiment. N.M.K. is grateful to CUI Hamburg for financial support. A.K.K. acknowledges the support from the Basque Government (grant IT-756-13 UPV/EHU) and from the Spanish Ministerio de Econom\'{i}a y Competitividad (grants FIS2016-76617-P and FIS2016-76471-P). This is a pre-print of an article published in Communications Physics. The final authenticated version
is available online at: \url{https://doi.org/10.1038/s42005-018-0032-5}
\end{acknowledgments}

\section*{Author contribution}
The experiment was coordinated by J.S., M.D., K.S. and P.W. Construction and set up of the experiment was done by P.W., B.R., T.K. and J.S.. Measurements and data analysis were performed by P.W. with support from B.R., T.K. and J.S. The theoretical calculations were performed by A.K.K. and N.M.K. The manuscript was written by P.W. with discussions involving all co-authors.

\section*{Additional information}

\subsection*{Competing financial interests}
The authors declare no competing financial interests.


\clearpage

\appendix

\onecolumngrid
\begin{center}
\textbf{\large Supplemental Material:\\Absolute strong-field ionization probabilities of ultracold rubidium atoms}
\end{center}
\twocolumngrid

\setcounter{figure}{0}
\setcounter{equation}{0}
\setcounter{table}{0}

\makeatletter
\renewcommand{\thefigure}{S\@arabic\c@figure}
\renewcommand{\theequation}{S\arabic{equation}}
\makeatother


\section*{Units}
\label{supp:units}

All equations are given in SI units where $\varepsilon_0 = 1 / (\mu_0 c^2)$ represents the electric constant defined by the speed of light in vacuum $c$ and the magnetic constant $\mu_0$. For conversion between atomic units (a.u.) and SI units these factors can be used:

\begin{alignat}{2}
 F_c &= \frac{E_h}{e a_0} \times 10^{-2} & &= 5.1422 \times 10^9 \: \text{V/cm},
 \label{eq:EauSI}\\
 I_c &= \frac{1}{2} \varepsilon_0 c F_c^2 & &= 3.50945 \times 10^{16} \: \text{W/cm}^2
 \label{eq:IauSI}
\end{alignat}

\noindent where $E_h$ is the Hartree energy, $e$ the elementary charge and $a_0$ the Bohr radius.

A conversion between the peak electric field $E_0^\text{SI}$ and peak intensity $I_0^\text{SI}$ in SI and atomic units ($F_0^\text{a.u.}$, $I_0^\text{a.u.}$) is then possible by applying the relations

\begin{align}
 E_0^\text{SI} &= F_c F_0^\text{a.u.}, & I_0^\text{SI} &= I_c I_0^\text{a.u.}.
 \label{eq:autoSI}
\end{align}

A conversion to the peak electric field strength in atomic units $F_0$ from a given peak intensity $I_0$ in SI units can be achieved by using the relation

\begin{equation}
 F_0 = \sqrt{\frac{I_0}{I_c}}.
 \label{eq:FauSI}
\end{equation}

\section*{Model ionization probabilities}
\label{supp:rates}

In order to quantify an ionization process, the ionization rate $w(t)$ is calculated as a measure of how many atoms per time-interval are ionized. For photoionization this rate depends on the instantaneous intensity $I$ of the light field which is time-dependent for a laser pulse. It is described by an oscillating field term and an intensity envelope $I(t) \Rightarrow w(I) = w(I(t)) \equiv w(t)$. When combining two ionization models, the total ionization rate

\begin{equation}
 w(t) = \max{\left\{ w_1(t), w_2(t) \right\}}%
 \label{eq:wmax}
\end{equation}

\noindent is determined by using the dominating ionization rate of the two models. Once the ionization rate $w(t)$ is known from the models presented below, the ionization probability $P(t)$, i.e. the fraction of ionized atoms with respect to the total initial number of atoms $N_i / N_0$, is determined by the following differential equation:

\begin{equation}
 \frac{\mathrm{d} P(t)}{\mathrm{d} t} = \left[ 1 - P(t) \right] w(t) .%
 \label{eq:pdgl}
\end{equation}

\noindent Solving the equation leads to the solution function

\begin{equation}
 P(t) = 1 - \exp{\left( - \int_{-\infty}^{t} \! w(t') \, \mathrm{d}t' \right)}%
 \label{eq:pion}
\end{equation}

\noindent that saturates at zero and unity for low and high ionization rates, respectively. Equation (\ref{eq:pion}) refers to the cumulated ionization probability at a specific time $t$ in the intensity envelope. In order to calculate the ionization probability after a single laser pulse

\begin{equation}
 P = \lim_{t \rightarrow \infty} P(t)%
 \label{eq:pfin}
\end{equation}

\noindent the limit for $t \rightarrow \infty$ has to be calculated. The ionization probabilities for rubidium using the models discussed below are presented in Fig.~3a and Fig.~4a in the main manuscript.

\subsection*{Multiphoton ionization}

In a multiphoton process the photon energy $E_\text{ph} = \hbar \omega$ is lower than the binding energy $E_I$ of the atoms so that the atom has to absorb multiple photons in order to release the bound electron into the continuum. The number of absorbed photons $m = \lceil E_I / E_\text{ph} \rceil$ refers to the order of the $m$-photon process and the ionization rate

\begin{equation}
 w_\text{MPI}(I) = \sigma_m \Phi^m(t) = \sigma_m \left[ \frac{I}{E_\text{ph}} \right]^m%
 \label{eq:wmpi}
\end{equation}

\noindent scales with the photon flux $\Phi(t) = I(t) / E_\text{ph}$ and thus with the laser intensity $I$ to the power of $m$ and is proportional to a multiphoton ionization cross section $\sigma_m$, generally expressed in units of cm$^{2m}$s$^{m-1}$.

The cross section depends on the wavelength and for a two-photon process in rubidium at 511~nm wavelength it has been measured and calculated to $\sigma_2 = 1.5 \times 10^{-49}$ cm$^4$ s \cite{Takekoshi:RbMPI}. At 1022~nm only a calculated five-photon cross section $\sigma_5 = 5 \times 10^{-139}$ cm$^{10}$ s$^4$ has been reported in the literature \cite{Edwards:MPI5Rb} that describes an ATI process. For reference, a calculated four-photon cross section for a slightly different wavelength at 1064~nm (Nd:YAG laser) $\sigma_4^N = 1.32 \times 10^{-107}$ cm$^8$ s$^3$ \cite{Delone:Rb4PhotonCS} is also available in the literature.

\subsection*{Tunnel ionization}

The tunnel rate of a bound electron inside an atom penetrating the barrier formed by the superposition of Coulomb potential and light field can be calculated using the Ammosov-Delone-Krainov (ADK) model \cite{PPT:Ionization, ADK:Tunnel, Spielmann:keVXray, Milosevic:StrongFieldAlkali} and the time-averaged rate for a linearly polarized laser pulse is obtained by

\begin{eqnarray}
 w_\text{ADK}(I) = \left| C_{n^*0} \right|^2 \sqrt{\frac{6}{\pi R(I)}} \frac{E_I}{\hbar}\nonumber\\
  \times R^{2 n^* - 1}(I) \exp{ \left( - \frac{R(I)}{3} \right)}%
 \label{eq:wadk}
\end{eqnarray}

\noindent where

\begin{equation}
 R(I) = \frac{4 E_I^\frac{3}{2}}{\hbar e} \sqrt{\frac{\varepsilon_0 c m_e}{I}}%
 \label{eq:R}
\end{equation}

\noindent and

\begin{equation}
 \left| C_{n^* 0} \right|^2 = \frac{2^{2n^*}}{n^* \Gamma{(n^* + 1)} \Gamma{(n^*)}}%
 \label{eq:Cn2}
\end{equation}

\noindent using the gamma function $\Gamma$ and the effective principal quantum number

\begin{equation}
 n^* = Z \sqrt{\frac{E_{I,H}}{E_I}}%
 \label{eq:nstar}
\end{equation}

\noindent given by the charge state $Z$ of the ion and the ionization potential $E_I$ related to the hydrogen ionization potential $E_{I,H} = 13.6$ eV.

Note that this is already the time-averaged result for an oscillating electric field of a laser pulse so that the total ionization rate can be obtained by integrating over the \emph{intensity-envelope} of the laser pulse. If an integration over the instantaneous intensity $I(t) = (c \varepsilon_0 / 2) \left| E(t) \right|^2$ of the field oscillations is performed, the prefactor $\sqrt{6 / [\pi R(I)]}$ has to be omitted in Eq.~(\ref{eq:wadk}).

\subsection*{Over-the-barrier ionization}

The over-the-barrier ionization intensity

\begin{equation}
 I_\text{OBI} = \frac{c}{128 \pi} \left( \frac{4 \pi \varepsilon_0}{e^2} \right)^3 \frac{E_I^4}{Z^2}%
 \label{eq:iobi}
\end{equation}

\noindent denotes the intensity where the electron can classically escape from the nucleus. Again, $E_I$ denotes the binding energy and $Z$ the charge state of the ion. In a simple threshold model, the ionization rate can be set to infinity once $I_\text{OBI}$ is exceeded such that ionization is certain and can be set to zero below this threshold:

\begin{equation}
 w_\text{OBI}(I) = \left\{
    \begin{array}{ll}
		 \infty, & I > I_\text{OBI}\\
          0, & I \leq I_\text{OBI}
		\end{array} \right. .%
 \label{eq:wobi}
\end{equation}

\end{document}